\documentclass[aps,prl,amssymb,showpacs]{revtex4}
\newcommand{\trl}{{\rm tr}\lambda}
\newcommand{\tru}{{\rm tr}u}
\newcommand{\trlto}{{\rm tr}(\lambda^2)}
\newcommand{\bsr}{\frac{{\rm Bulk\ elastic\ energy}}{{\rm Shear\ elastic\ energy}}}
\newcommand{\utr}{u_{ik}^{\perp}}
\newcommand{\lxx}{(\lambda xx)}
\usepackage{graphicx}
\begin{document}
\title{Point defect in solids: Shear dominance of the far-field energy}
\author{Jeppe C. Dyre} 
\affiliation{DNRF centre of viscous liquid dynamics ``Glass and time,''
IMFUFA, Building 27, Department of Sciences,
Roskilde University, Postbox 260, DK-4000 Roskilde, DENMARK}
\date{\today}

\begin{abstract} 
It is shown that the elastic energy far from a point defect in an isotropic solid is mainly shear elastic energy. The calculation, which is based on a standard dipole expansion, shows that no matter how large or small the bulk modulus is compared to the shear modulus, less than 10\% of the distant point defect energy is associated with volume changes.
\end{abstract}
\maketitle

An isotropic solid has two independent elastic constants, the shear modulus $G$ and the bulk modulus $K$. It has been suggested that when a physical property depends on both $G$ and $K$, the dependence on the shear modulus is often the most important \cite{dyr04}. Examples of this ``shear dominance'' were given by Granato in his important paper from 1992 \cite{gra92}; they include the fact that defect energies vary only slightly with the bulk modulus, but are linearly dependent on the shear modulus (for Cu he estimated that only about 3{\%} of the energy of an interstitial is bulk elastic energy). In Ref. \cite{dyr04}, which dealt with the non-Arrhenius viscosity of viscous liquids and its possible explanation in terms of temperature-dependent instantaneous elastic constants \cite{dyr06}, the following result was proved: If temperature dependence is quantified in terms of log-log derivatives, at least 92{\%} of the temperature dependence of the molecular vibrational mean-square displacement over temperature comes from the instantaneous shear modulus, whereas at most 8{\%} is due to that of the instantaneous bulk modulus. 

How general is ``shear dominance''? Is it coincidental, or is it a fairly general theme of three-dimensional elasticity? Of course, a property like the Poisson ratio depends equally on the bulk and shear moduli, so shear dominance cannot be a completely general phenomenon. Nevertheless, there are several examples of it so it deserves to be investigated. As one contribution to shedding light on this question, we below calculate the maximum ratio of bulk elastic energy to shear elastic energy far from an arbitrary point defect in an isotropic solid. Based on a standard dipole expansion \cite{mur82} it is shown that less than 10\% of the distant elastic energy is bulk elastic energy, i.e., associated with density changes. 

Although our main objective is to identify the dominant contribution to the elastic energy far from a point defect, it should be noted that according to Zener's ``strain energy model'' \cite{zen42} most energy associated with a point defect is elastic. Elastic energies of point defects fall of rapidly (as $r^{-6}$), as evident from Eshelby's famous solution of  the problem of an elipsoidal inclusion \cite{esh57,kie97,ona00,kie02,li04,mat05,fis06}, a result which is general. This means, of course, that most of the elastic energy is located close to the point defect and that a dipole expansion is not realistic for calculating the total defect energy. Nevertheless, our results may be taken as an indication of what contributes most to the total defect energy, the shear or the bulk elastic energies in the defect surroundings. 

The nature of the far-field deformation is also important for understanding and modeling the long-range properties of defect-defect interactions. If, for instance, the displacement field is dominated by shear displacements, a defect will only interact weakly with one far away that is dilational close to its center. Some time ago Andreev discussed the unique topological characteristics of point defects in three dimensions \cite{and95}. A comprehensive review of point defect properties focusing on the vacancy was given by Kraftmakher \cite{kra98}. The question of the nature of the deformation far from a point defect is relevant also for applications in materials science, e.g., for understanding fracture \cite{kie02}. In its macroscopic description the question of the far-field properties of a point defect has been studied because of its important for understanding the mechanical properties of polycrystals and composites \cite{mur82,esh57,esh61}. In this contex Onaka recently calculated the elastic shear energy due to a macroscopic point defect (an ``inclusion'') \cite{ona05}, but did not study the general case, leaving undetermined what is our focus here, the ratio between shear and bulk energy (see also Ref. \cite{fis06}). Garikipati and co-workers \cite{gar05} recently discussed the role of continuum elasticity in determining the formation energy of a point defect utilizing Eshelby's result for the work done by an external stress during the transformation of an inclusion \cite{esh57}, an interesting paper that also elicidates the limitations of this approach as compared to atomistic calculations. We mention these works also to emphasize the continuing interest in the far-field properties of point defects and their macroscopic analogs.

First some preliminaries: We use the standard Einstein summation convention. The point defect is located at the origin, is modelled as follows: Imagine a small sphere of radius $R$ surrounding the point defect with all atoms within the sphere removed. The effect of the defect on the surroundings are represented by a suitable distribution of external forces $F_i$ applied to the surface of the sphere. We define a tensor $\lambda_{ij}$ as the following integral over the surface of the sphere (where $dA$ is the area element):

\begin{equation}\label{1}
\lambda_{ij} = \oint_{|{\bf y}|=R}  F_i({\bf y})\ y_j\,dA\,.
\end{equation}
Because the force distribution models the effect of the atoms within the sphere on the surroundings, the angular momentum of the force distribution must be zero. This implies that $\lambda$ is symmetric: $\lambda_{ij}=\lambda_{ji}$. 

Sharp brackets $\langle\rangle_r$, referred to as ``averages,'' denote integrations over the surface of a sphere with radius $r\gg R$ centred at the origin. The following identities become useful later on:

\begin{equation}\label{2}
\langle x_i x_j\rangle_r\, =\,
\frac{r^2}{3}\delta_{ij}\,
\end{equation}
and

\begin{equation}\label{3}
\langle x_i x_j x_k x_l\rangle_r\, =\,
\frac{r^4}{15}\ \Big( \delta_{ij}\delta_{kl} + \delta_{ik}\delta_{jl} + \delta_{il}\delta_{jk} \Big)\,.
\end{equation}
If $\lxx$ is an abbreviation of $\lambda_{ij}x_ix_j$, we find from these identities that

\begin{equation}\label{4}
\langle\lxx\rangle_r \,=\,
\frac{r^2}{3}\trl\,
\end{equation}
and

\begin{equation}\label{5}
\langle\lxx^2\rangle_r\,=\,
\frac{r^4}{15}\Big((\trl)^2+2\trlto\Big)\,.
\end{equation}
Moreover, if $\lambda x$ is the vector whose i'th component is $\lambda_{ij}x_j$, we have

\begin{equation}\label{6}
\langle(\lambda^2xx)\rangle_r\,=\, 
\langle(\lambda x)^2\rangle_r\,=\,
\frac{r^2}{3}\trlto\,.
\end{equation}

As mentioned, far from the defect both the bulk and the shear elastic energies vary with the distance from the origin as $r^{-6}$. When one averages over a sphere with radius $r\gg R$, the result for both the bulk and the shear elastic energy must be a scalar varying with distance as $r^{-6}$ that is of second order in the forces $F_i$. It follows from the below calculation that these two scalars are both uniquely determined by $G$, $K$, and the $\lambda$-matrix. Consequently, because these two scalar functions are of second order in the forces, the ratio of bulk to shear elastic energy must have the following general structure:

\begin{equation}\label{7}
\bsr =
\frac{A (\trl)^2 + B \trlto}{C(\trl)^2 +  D\trlto}\,.
\end{equation}
Defining $\alpha=\trlto/(\trl)^2$ this ratio is $(A+B\alpha)/(C+D\alpha)$. The quantity $\alpha$ varies between $1/3$ and $\infty$: By normalizing we may assume that $\trl=1$; if the eigenvalues are denoted by $\mu_i$ we thus have $\mu_1+\mu_2+ \mu_3=1$ and consequently $\alpha=\mu_1^2+\mu_2^2+\mu_3^2$. From this it follows that $\alpha$ varies between $1/3$ and $\infty$. Since the energy ratio of Eq. (\ref{7}) is a monotonous function of $\alpha$ and since perfect isotropy ($\alpha=1/3$) implies zero bulk elastic energy (in this case the displacement field is radially symmetric varying as $r^{-2}$ which implies no volume changes), the maximum bulk to shear elastic energy ratio arises in the limit $\alpha\rightarrow\infty$. Thus in the below calculation we may ignore all terms with $\trl$ and keep only terms with a $\trlto$ factor. 

Poisson's ratio $\sigma$ is defined \cite{lan70} by

\begin{equation}\label{8}
\sigma =
\frac{1}{2}\ \frac{3K - 2G}{3K + G}\,.
\end{equation}
If a force ${\bf F}$ is applied at the origin of an isotropic elastic continuum, the displacement field at the point $(x',y',z')$ is given \cite{lan70} by 

\begin{equation}\label{9}
{\bf u}(x',y',z')\, \propto\,
\frac{(3-4\sigma){\bf F} + ({\bf F\cdot n}'){\bf n}'}{r'}\,,
\end{equation}
where $r'^2=x'^2+y'^2+z'^2$ and ${\bf n}'=(x',y',z')/r'$ is the unit vector pointing from the origin to $(x',y',z')$. It is convenient to introduce the variable

\begin{equation}\label{10}
\Lambda \,\equiv\,
2-4\sigma\,,
\end{equation}
in terms of which Eq. (\ref{9}) becomes

\begin{equation}\label{11}
u_i \,\propto\,
(\Lambda +1)\frac{F_i}{r'} + \frac{F_jn'_jn'_i}{r'}\,.
\end{equation}

We proceed to perform a standard dipole expansion by first noting that, if ${\bf y}$ is the coordinate for a point on the small sphere surrounding the defect ($|{\bf y}|=R$) and ${\bf x}$ is the coordinate for the point of interest far away, to lowest order in $|{\bf y}|/|{\bf x}|=R/|{\bf x}|$ we have if $r\equiv|{\bf x}|$

\begin{equation}\label{12}
\frac{1}{|{\bf x}-{\bf y}|} =
\left({\bf x}^2+{\bf y}^2-2{\bf x\cdot y}  \right)^{-1/2} = 
r^{-1}\left(1-2\frac{{\bf x\cdot y}}{r^2}\right)^{-1/2} =
r^{-1} + \frac{{\bf x\cdot y}}{r^3}\,.
\end{equation}
Similarly 

\begin{equation}\label{13}
\frac{1}{|{\bf x}-{\bf y}|^3} =
\left({\bf x}^2+{\bf y}^2-2{\bf x\cdot y}  \right)^{-3/2} = 
r^{-3}\left(1-2\frac{{\bf x\cdot y}}{r^2}\right)^{-3/2} =
r^{-3} + 3\frac{{\bf x\cdot y}}{r^5}\,.
\end{equation}
To calculate the displacement field at point $\bf x$ we first note that when the force $F_i({\bf y})$ is integrated over the small sphere radius $R$, the result is zero. Thus when one integrates over the small sphere, the first term of Eq. (\ref{11}) to lowest order in $1/r$ becomes

\begin{eqnarray}\label{14}
& (\Lambda +1) & \oint_{|{\bf y}|= R}  \frac{F_i({\bf y})}{|{\bf x}-{\bf y}|} \,dA
=(\Lambda +1)
\oint_{|{\bf y}|=\rm R}  F_i({\bf y}) \frac{{\bf x\cdot y}}{r^3} \,dA
\nonumber\\
& = & 
(\Lambda +1)\ r^{-3}\lambda_{ij}x_j\,.
\end{eqnarray}
Similarly, to lowest order the second term of Eq. (\ref{11}) gives the following contribution to the displacement field

\begin{eqnarray}\label{15}
\oint_{|{\bf y}|=R}
\frac{F_j({\bf y})(x_j-y_j)(x_i-y_i)}{|{\bf x}-{\bf y}|^3}\,dA &  = & 
\oint_{|{\bf y}|=R} F_j({\bf y}) \left(r^{-3} + 3\frac{{\bf x\cdot y}}{r^5}\right)
(x_j-y_j)(x_i-y_i)\,dA\nonumber \\
& = &
- r^{-3}\lambda_{ji}x_j 
-r^{-3}\lambda_{jj}x_i 
+ 3 r^{-5}\lambda_{jl}x_jx_lx_i\,.
\end{eqnarray}
Thus, if proportionality is replaced by equality for simplicity of notation -- which is OK because we only wish to calculate an energy ratio and have already dropped the overall proportionality constant of Eq. (\ref{9}) -- we find

\begin{equation}\label{16}u_ i =
\Lambda r^{-3}\lambda_{ij}x_j - r^{-3}\trl\ x_i 
+ 3 r^{-5}\lambda_{jl}x_jx_lx_i\,.
\end{equation}

Next we calculate the stain tensor. First, note that if $\partial_k$ is the partial derivative with respect to $x_k$, we have $\partial_k r^{-n} = (-n) r^{-(n+2)}x_k$. Thus

\begin{eqnarray}\label{17}
\partial_k u_i & = &
-3\Lambda r^{-5}\lambda_{ij}x_jx_k 
+\Lambda r^{-3}\lambda_{ij}\delta_ {jk}
+3r^{-5}\trl\, x_ix_k
-r^{-3}\trl\, \delta_{ik}
-15r^{-7}\lambda_{jl}x_jx_lx_ix_k\nonumber\\
& + & 3r^{-5}\lambda_{jl}\delta_{jk}x_lx_i
+3r^{-5}\lambda_{jl}x_j\delta_{lk}x_i
+3r^{-5}\lambda_{jl} x_jx_l\delta_{ik}\,.
\end{eqnarray}
All terms in this expression vary with $r$ as $r^{-3}$. Consequently, this term is common to the bulk and shear elastic energies and may be dropped from our calculation of their ratio. The calculation is simplified notationally by putting $r=1$ (or, equivalently, replacing $x_i$ by $x_i/r$). When this convention is adopted, the strain tensor is given by (ignoring the factor $2$ in the strain tensor definition, $u_{ik}=(\partial_i u_k +\partial_k u_i )/2$)

\begin{eqnarray}\label{18}
u_{ik} &  = &
2\left(- \trl  +3\lxx\right)\delta_{ik} 
+2\Lambda\lambda_{ik}
+6\trl x_ix_k
-30\lxx x_ix_k\nonumber\\
& - & 3\Lambda (\lambda_{ij}x_jx_k + \lambda_{kj}x_jx_i)
+6 \lambda_{kj}x_jx_i
+6 \lambda_{ij}x_jx_k
\nonumber\\
&  = & 2 \left(- \trl +3\lxx\right)\delta_{ik}
+2\Lambda\lambda_{ik}
+\left(6\trl -30\lxx\right) x_ix_k\nonumber\\
& +& (6-3\Lambda) 
\left(\lambda_{ij}x_jx_k + \lambda_{kj}x_jx_i \right)
\end{eqnarray}
Throwing out terms with $\trl$ we end with the following expression for the strain tensor

\begin{equation}\label{19}
u_{ik} =
6\lxx\delta_{ik}
+2\Lambda\lambda_{ik}
-30\lxx x_ix_k
+(6-3\Lambda) \left(\lambda_{ij}x_jx_k
+ \lambda_{kj}x_jx_i\right)\,.
\end{equation}

To calculate the bulk elastic energy we need the trace of this which, when again terms with $\trl$ are dropped, is given as ($x_ix_i=r^2$ is put equal to unity)

\begin{equation}\label{20}
\tru =
18\lxx -30 \lxx +(6-3\Lambda)2\lxx =
-6\Lambda \lxx\,.
\end{equation}
The bulk elastic energy density \cite{lan70} averaged over the sphere with radius $r$ (subsequently put equal to unity) is, when use is made of Eq. (\ref{5}) and terms with $\trl$ are ignored, given by

\begin{equation}\label{21}
{\rm Bulk\ elastic\ energy} = \frac{K}{2}\ \langle({\rm tr} (u))^2 \rangle_r=
\frac{K}{2}\ 36\Lambda^2\langle \lxx^2\rangle_r=
\frac{12}{5}K\Lambda^2\trlto\,.
\end{equation}

To find the shear elastic energy we need the transverse part of the strain tensor, $\utr$ (in terms of which the shear energy density is $G\utr\utr$) defined as \cite{lan70}:
\cite{lan70}

\begin{eqnarray}\label{22}
\utr & \equiv &
u_{ik}-\frac{1}{3}{\rm tr}(u)\delta_{ik}\nonumber\\
& = &
2(3+\Lambda)\lxx\delta_{ik} +2\Lambda\lambda_{ik}
-30\lxx x_ix_k+3(2-\Lambda)
\left( \lambda_{ij}x_jx_k+\lambda_{kj}x_jx_i\right)\,.
\end{eqnarray}
Squaring and summing of all the elements of the transverse strain tensor, which is required to calculate the shear elastic energy, leads to the following ($x_ix_i=r^2$ is again put equal to unity):

\begin{eqnarray}\label{23}
\utr\utr & = & 
4(3+\Lambda)^2\lxx^2 3 
+4\Lambda^2\trlto
+900\lxx^2
+9(2-\Lambda)^2[2(\lambda x)^2+2\lxx^2]\nonumber\\
& + & 8\Lambda(3+\Lambda)\lxx\trl
-120(3+\Lambda)\lxx^2
+12(3+\Lambda)(2-\Lambda)2\lxx^2
-120\Lambda\lxx^2\nonumber\\
& + & 12\Lambda(2-\Lambda)2(\lambda^2 xx)
-180(2-\Lambda)\lxx 2\lxx\,.
\end{eqnarray}
Averaging this expression (ignoring all $\trl$-terms) leads to:

\begin{eqnarray}\label{24}
\langle\utr\utr\rangle_r & = & \Big(
12(3+\Lambda)^2\frac{2}{15} + 4 \Lambda^2+900 \frac{2}{15}
+18(2-\Lambda)^2(\frac{1}{3}+\frac{2}{15})\nonumber\\
& - & 120(3+\Lambda) \frac{2}{15} 
+24(3+\Lambda)(2-\Lambda) \frac{2}{15}-120\Lambda \frac{2}{15}\nonumber\\
& + &24\Lambda(2-\Lambda)\frac{1}{3}-360(2-\Lambda) \frac{2}{15}
\Big) \trlto
\nonumber\\
& = & \frac{2}{5}\left(7\Lambda^2+12\Lambda+108\right)\trlto\,.
\end{eqnarray}
Summarizing, we find that

\begin{equation}\label{25}
\bsr\le
\frac{(12/5)K\Lambda^2}{(2/5)G(7\Lambda^2+12\Lambda+108)}=
\frac{K}{G}\ 
\frac{6\Lambda^2}{7\Lambda^2+12\Lambda+108}\,.
\end{equation}
In terms of the dimensionless variable

\begin{equation}\label{26}
k  \equiv \frac{K}{G}\,,
\end{equation}
we have $\Lambda=6/(3k+1)$ which, when substituted into Eq. (\ref{24}), leads to

\begin{equation}\label{27}
\bsr\le
 \frac{2k}{9k^2+8k+4}\,.
\end{equation}
The derivative of the fraction on the right with respect to $k$ is zero when $2(9k^2+8k+4)=2k(18k+8)$, implying $k=2/3$. Thus the maximum bulk elastic energy is when $k=2/3$. In conclusion, 

\begin{equation}\label{28}
\bsr\le
\bsr \Big(k=\frac{2}{3}\Big) \le
\frac{1}{10}\,.
\end{equation}
More typically, $k=5/2$ leads to a maximum bulk shear ratio of $20/337$ which is roughly 6\%.

One important unsolved problem of condensed-matter physics is the origin of the non-Arrhenius average relaxation time of glass-forming liquids, where in most cases one observes an activation energy that increases quite a lot upon cooling. One class of theories are the elastic models (recently reviewed in \cite{dyr06}) according to which the activation energy is some linear combination of the instantaneous bulk and shear moduli. As the regards the temperature dependence of the activation energy the shear modulus completely dominates (contributing at least 90\% \cite{dyr04}). A so-called flow event -- the jump in configuration space from one to another potential energy minumum -- is usually localized in real space. Since the surrounding ultraviscous liquid on the short time scale may be regarded as a solid, the molecular displacements induced by a flow event may be regarded as those of a point defect (in a disordered solid, albeit). The above result sheds light on the nature of far-field flow-event induced displacements by emphasizing ``shear dominance'' -- that the shear modulus is much more important than the bulk.

In an interesting book from 1986 Varotsos and Alexopoulos (VA) discussed point defect energies in solids, as well as activation energies for point defect diffusion \cite{var86}. VA concluded that these energies can always be written as the (isothermal) bulk modulus times some microscopic volume. Thus point defect energies and activation energies always scale with the bulk modulus. This result is at variance with the above calculation. In practical terms, of course, Poisson's ratio usually does not vary very much, so the bulk and shear moduli are roughly proportional. This means that in many cases it is difficult to determine whether the bulk or the shear modulus controls things. Admittedly, VA explicitly argue in their book that it {\it is} the bulk, and not the shear modulus, which is important, but they do not base this on calculations similar to ours (but, e.g., thermodynamic arguments). We do not claim that the above calculation disproves VA, but it may encourage other researchers to look into the question. Surely, the idea that many important physical properties are controlled by as simple quantities as the elastic moduli deserves to be investigated fully.

\acknowledgments
This work was supported by the Danish National Research Foundation's Centre for Viscous Liquid Dynamics ``Glass and Time.''

\end{document}